\documentclass[12pt]{article}
\usepackage{cospar}
\usepackage{natbib}
\usepackage{url}

\usepackage{graphicx}
\usepackage[figuresright]{rotating}

\newcommand\apj{Astrophys. J.}% 
\newcommand\apjl{{Astrophys. J. Lett.}}% 
\newcommand\aap{{A\&A}}% 
\newcommand\mnras{{MNRAS}}% 
\newcommand\ssr{{Space~Sci.~Rev.}}% 
\newcommand\jgr{{J.~Geophys.~Res.}}% 
\def\HI{\hbox{H$^{\rm o}$}}
\def\OI{\hbox{O$^{\rm o}$}}
\def\HeI{\hbox{He$^{\rm o}$}}
\def\HII{\hbox{H$^{\rm +}$}}
\def\ArI{\hbox{Ar$^{\rm 0}$}}
\def\Rhp{\hbox{$R_{\rm HP}$}}
\def\deeg{\hbox{$^{\rm o}$}}
\def\kms{\hbox{km s$^{\rm -1}$}}
\def\cc{\hbox{cm$^{\rm -3}$}}
\def\cmtwo{\hbox{cm$^{\rm -2}$}}
\def\Rgd{\hbox{$R_{\rm g/d}$}}

\title{WHY STUDY INTERSTELLAR MATTER VERY CLOSE TO THE SUN?} 

\author{P.C. Frisch\address{University of Chicago, Department of Astronomy
	and Astrophysics, 5640 South Ellis Avenue, Chicago, IL  60637 USA}}

\begin{document}
% typeset front matter
\maketitle

\begin{abstract}
Interstellar matter (ISM) sets the boundary conditions of the
heliosphere and dominates the interplanetary medium.  The heliosphere
configuration has varied over recent history, as the Sun emerged from
the Local Bubble and entered a turbulent outflow of cloudlets
originating from the Scorpius-Centaurus Association direction.
Several indicators suggest that the local interstellar magnetic field
is weak and parallel to the galactic plane.  Observations of the
interaction products of the ISM with the heliosphere, such as pickup
ions and anomalous cosmic rays, when combined with data on the ISM
towards nearby stars, provide unique constraints on the composition
and physical properties of nearby gas.  These data suggest abundances
in nearby ISM are subsolar, and that gas and dust are not well mixed
at the solar location.
\end{abstract}
%%=======================================================================
\section*{INTRODUCTION}

Interstellar gas and dust within $\sim$30 pc of the Sun provide a
unique opportunity to evaluate the chemical composition and physical
properties of diffuse interstellar clouds in a small region of space.
Several motivations for studying local interstellar matter (ISM, LISM)
stand out.
\begin{enumerate}
\item The ISM sets the boundary conditions of the heliosphere.  By
analogy, the ISM also sets the boundary conditions for the
astrospheres of extrasolar planetary systems.  Through the
heliosphere-cosmic ray flux (and by analogy the astrosphere-cosmic ray
flux) connection, the distribution of stable planetary climates may,
in part, depend on the present and historical physical properties of
the ISM surrounding each system.
\item Although many details of the chemical composition of the ISM are
known, the overall picture is poorly understood because elements are
observed only in the gas phase of the ISM, line blending in distant
stars confuses the interpretation of data, \HII\ is unobserved in low
column density sightlines, and the interstellar grain composition,
gas-dust mixing and chemical homogeneity are unknown.
\item 
The configuration of the heliosphere depends on the physical
properties of the surrounding ISM.  The Sun moves through space, and
currently the Sun is embedded in an outflow of material from the
Scorpius constellation.  As the Sun travels through space, the
heliosphere is responsive to ``galactic weather'' patterns.  Also, the
plane of the ecliptic is tilted by $\sim$60\deeg\ with respect to the
galactic plane, introducing a north/south asymmetry in the heliosphere
because the local interstellar magnetic field appears to be parallel to the
galactic plane.
\item Turbulence in the ISM, whether macroscopic or microscopic, is
poorly understood.  Interstellar absorption lines at the velocity of
the cloud surrounding the solar system (also known as the Local
Interstellar Cloud, LIC) are not observed towards the nearest star,
$\alpha$ Cen ($\sim$45\deeg\ from the nose of the heliosphere),
suggesting either a turbulence-driven velocity discontinuity, or a
physically distinct interstellar cloudlet in this direction.  The
heliosphere responds differently to the case where velocity
discontinuities, versus ``empty'' space, distinguish cloudlets.
\item 
Lastly, dust and neutral gas from the LIC enter the solar system,
where observations of the interaction products directly sample the
chemical composition and ionization levels of the parent ISM.  The
combination of ISM observations towards nearby stars and $in~situ$
observations of He, H, pickup ions, and anomalous cosmic rays in the
heliosphere provides strong constraints on the physical conditions of
the surrounding cloud.

\end{enumerate}
These unique qualities of the ISM close to the Sun offer a space
laboratory for studying cosmic processes that is not duplicated
elsewhere in the Universe, and are elaborated on below.

\section*{BOUNDARY CONDITIONS OF HELIOSPHERE AND ASTROSPHERES}

The first spectral observations of the interplanetary Lyman-$\alpha$
glow established that neutral gas at interstellar velocities has
penetrated the heliosphere (Adams \& Frisch 1977).  The heliosphere
responds to the relative Sun-cloud velocity and the partial pressure
of the constituents of the LIC.  As the Sun journeys through the Milky
Way Galaxy, the heliosphere configuration varies in response to the
properties of the surrounding ISM (Frisch 1993, 1997; Zank \& Frisch
1999; Mueller et al.\ 2001, 2002).  Using recent Hipparcos results for
the solar apex motion (Dehnen \& Binney 1998), the Sun moves through
the local standard of rest at $\sim$14 parsecs per million years, and
has a relative velocity with respect to the LIC of $\sim$26 parsecs
per million years.
%%\citep{AdamsFrisch:1977}
%%\citep{Frisch:1993a,Frisch:1997,ZankFrisch:1999,Muelleretal:2001,Muelleretal:2002, Frischetal:2002b}

The Sun is currently immersed in an outflow of ISM from the region of
the Scorpius-Centaurus Association (see below), and the distance to
the outflow surface in the heliosphere tail direction is $\sim$0.9 pc.
Hence the Sun is likely to have entered this outflow within the past
10$^5$ years, and possibly within the past $\sim 10^4$ years depending
on cloud geometry (Frisch 1994).  Prior to that the Sun was immersed
in the Local Bubble interior, a region of low density hot plasma
($\sim$0.005 cm$^{-3}$, 10$^6$ K) where the heliosphere was probably
large (heliopause radius in nose direction, \Rhp$\sim$300 AU, Frisch et al. 2003, FMZL03).  
An encounter with a diffuse cloud of somewhat higher density,
$n$(\HI)$\sim$10 cm$^{-3}$, would contract the heliosphere
(\Rhp$\sim$15 AU) with dramatic consequences for galactic cosmic ray
modulation, and cosmic ray fluxes at the Earth's surface.
In contrast, todays heliosphere has \Rhp$\sim$100 AU (see
other papers in this issue).  Fig. 1 illustrates the motion of the Sun
through the Local Bubble and with respect to the Cluster of Local 
Interstellar Clouds (CLIC).
%%\citep{Frisch:1994}
%%\citep{Frischetal:2002}
%%\citep{Frischetal:2002b}

\begin{figure}[ht]
\vspace*{3.9in}
%\special{psfile=berry.eps voffset=-107 vscale=60 hscale=60  hoffset=75 }
$~$ \\
$~$ \\
\caption{
ISM within 300 pc of Sun.  Figure based on E(B-V) maps
of Lucke (1978) and molecular clouds as shown in CO maps of Dame et al. (1987).
The three subgroups of the Scorpius-Centaurus Association (SCA) 
(Upper Scorpius, Upper Centaurus-Lupus, Lower Centaurus-Crux) are shown,
as is the location of the Geminga pulsar (towards Orion).  The yellow arrow
indicates the solar apex motion.  The dark blue arrow indicates the bulk
flow vector of the CLIC in the LSR.  The light blue arrow shows the
LIC LSR velocity.  
\label{densities}}
\end{figure}

%%\citep{Dame:1987,Lucke:1978}
%%\citep{Roble:1991,Tinsley:2000}
%%\citep{MarshSvensmark:2000,Frischetal:2002b}

The influence of the galactic cosmic ray flux on the global electrical
circuit (Roble 1991,Tinsley 2000) provides a tentative link between
heliosphere dimensions and the 1 AU cosmic ray flux.  A positive
correlation between low altitude tropospheric cloud cover ($\le$3 km
altitude) and neutron monitor counts (a direct tracer of $\ge$0.5 GeV
cosmic ray fluxes at Earth, FMZL03) has been established
over one solar cycle ($\sim$1985--1995, Marsh and Svensmark 2000).
Evidently cosmic rays interacting with the Earth's atmosphere produce
aerosols which seed water vapor condensation and cloud formation.
During the period of time the Sun was immersed in the Local Bubble
interior (probably for many millions of years), the heliosphere should
have been large with maximum galactic cosmic ray
modulation, and in the absence of surrounding ISM neutrals anomalous
cosmic rays would have been minimal.  If further investigation
confirms the correlation between low altitude cloud cover and the 1 AU
cosmic ray flux, then it would appear that heliospheric modulation of
the galactic and anomalous cosmic ray fluxes has a direct influence on
terrestrial climate, and a large heliosphere may be a factor in
climate stability.

By analogy, the ISM also sets the boundary conditions for the
astrospheres of extrasolar planetary systems (Frisch 1993), although
variations in stellar winds must also be considered (Wood et
al. 2001).  Through the heliosphere-cosmic ray (and by analogy
astrosphere-cosmic ray) connection, the physical properties of the
present and historical ISM surrounding each system may be a factor in
the distribution of stable planetary climates.  Although Earth-mass
planets have not yet been discovered, over 10\% of the more than 100
planets now discovered have semimajor axes near $\sim$1 AU (for a list
of discovered planets, see URL http://exoplanets.org).  Spectral types
for the central star are typically F, G, or K, and include both giant
and main sequence objects.  Mass-loss characteristics of cool stars
vary with stellar activity levels, which are traced by X-ray surface
flux, and decrease with stellar age (Wood et al. 2002).  The
astrospheres of extrasolar planetary systems may thus vary
substantially from the heliosphere, both because of the surrounding
ISM and stellar activity levels.  These considerations will be
important in evaluating the yet-to-be-discovered climates of
extrasolar planets, as well as for understanding the climatic history
of the Earth.
%%\citep{Frisch:1993a}
%%\citep{Woodetal:2001}
%%\citep{Woodetal:2002}

\section*{CHEMICAL COMPOSITION, PHYSICAL PROPERTIES, AND GAS-TO-DUST MASS RATIO
OF SURROUNDING CLOUD }

The low opacity of the CLIC to the cloud surface in the heliosphere
tail direction yields relatively high average ionizations of \HI\ and
\HeI\ at the solar location.  Observations of the ISM towards more
distant stars yields sight-line averaged abundances which may be
misleading in the presence of ionization gradients.  \HII\ is
generally unobserved in low column-density sightlines.  Both \HI\ and
\HII\ must be included for determining the chemical composition of the
LIC, since most of the observed ions are present in both neutral and
ionized regions (i.e. He, Ne, Ar, and elements with first ionization
potentials less than 13.6 eV).  The correct gas-phase abundances are
required to evaluate gas-dust mixing and the chemical homogeneity of
the ISM.

The LIC is unique in that radiative transfer models are constrained by
both observations of the ISM towards nearby stars, and by $in~situ$
observations of ISM interaction products inside of the solar system
(Slavin \& Frisch 2002, SF02).  The ISM composition and ionization at
the solar location has been studied using the radiative
characteristics of the material, with the models constrained by
observations of ISM towards $\epsilon$ CMa (using data of Gry \&
Jenkins 2001) and by pickup ion data for an assumed $n$(\HeI)=0.016$\pm$0.002 \cc\
in the LIC.  Within the past year new data
offers an improved basis for selecting among the 25 models generated
by the SF02 study.  Studies of \OI/\HI\ in the generic ISM now shows
that the ratio is relatively constant and $\sim$410 (Andre et
al. 2002).  Voyager observations of anomalous cosmic rays yield data
on interstellar argon (Cummings \& Stone 2002) which is partially
ionized in local ISM (\ArI/Ar$\sim$0.20--0.24).  More extensive data on
pickup ions are also available (Gloeckler \& Geiss 2002), as
well as better $in~situ$ observations of \HeI\ (Witte et al. 1996, and
private communication).  Two models provide the best match to the
range of the available data for an assumed $n$(\HeI)=0.016$\pm$0.002 \cc.  These two best models are Models 2 and
18, and the parameters for these models are shown in Table 1.  These
models, based on the assumption that O/H$\sim$400 PPM in d$<$3 pc ISM
(applying results of Andre et Andre et al. 2002 to the LISM), show that N and S are present at $\sim$60\% of
solar abundance while C is mildly subsolar (SF02).  For this
comparison, solar abundances from Holweger (2001) are assumed.
Alternatively, assuming O and C solar abundances from Allende Prieto et
al. (2001) would imply unrealistic supersolar C abundances.
%%\citep{SlavinFrisch:2002}
%%\citep{Holweger:2001,Prietoetal:2001}
%%\citep{Witte:1996}
%%\citep{GryJenkins:2001}
%%\citep{AndreHowketal:2002}
%%\citep{CummingsStone:2002} 
%%\citep{GloecklerGeiss:2002}

%%%%%%%%%%%%%%%%%%%%%%%%%%%%%%%%%%%%
%%\input{table.tex}
\begin{table}[th]
\vspace{-8mm}
\begin{minipage}{90mm}
  \caption{LIC Parameters for Best Models$^{a}$}
\begin{tabular}{lcc}
{Quantity} & {Model 2} &{Model 18}  \\
\hline
\\
\underline{Assumed Parameters:} & & \\
%%$n_{\rm \HI+\HII}$ (\cc)& 0.273 & 0.300 \\
$n_{\rm tot}$ (\cc)& 0.273 & 0.300 \\
log $T_{\rm h}$ (K) & 6.0  & 6.1 \\
$B_{\rm o}$ ($\mu$G) & 5.0 & 3.0 \\
$N_{\rm H^o} $ (10$^{17}$ \cmtwo)$^{b}$ & 6.5 & 6.5 \\
\\
\underline{ Results for ISM at Solar~Location: } & & \\
$n$(\HI) (\cc) & 0.208 & 0.242 \\
$n$(\HeI) (\cc) & 0.015 & 0.017 \\
$n$(e) (\cc) & 0.098 & 0.089 \\
$\chi$(H)$^{c}$ & 0.287 & 0.234 \\
$\chi$(He)$^{c}$ & 0.471 & 0.448 \\
log $N_{\rm tot}$ (\cmtwo)$^{b}$ & 18.03 & 17.98 \\
$N$(\HI)/$N$(\HeI)$^{b}$  & 11.6 & 12.1 \\
T (K) & 8,230 & 8,140 \\
\hline
\end{tabular}
$^{a}$ {Models are from Slavin \& Frisch (2002) and Frisch \& Slavin (2003).
$n_{\rm tot} = n$(\HI)+$n$(\HII), 
while log $N_{\rm tot} = N$(\HI)+$N$(\HII).}
$^{b}$ {The column density $N$(H$^o$) is the \HI\ column density 
between the Sun and cloud surface, etc.
In this case the cloud is the
sum of the two interstellar cloudlets within $\sim$3 pc of the Sun 
in the $\epsilon$ CMa direction (Gry \& Jenkins 2001).}
$^{c}$ {$\chi$(H,He) is the ionization fraction of element H,He.} 
\end{minipage}
\end{table}

%%%%%%%%%%%%%%%%%%%%%%%%%%%%%%%%%%%%

The filtration of neutrals in the heliosheath regions is a key unknown
quantity.  For example, filtration factors for Ar range from
$\sim$0.64 (Cummings \& Stone 2002) to $\sim$0.99 (Mueller and Zank
2002).  Hydrogen filtration factors are predicted on a purely
theoretical basis (e.g. Ripken \& Fahr 1983, Izmodenov et al. 1999),
however the radiative transfer models also yield an estimate of
hydrogen filtration in the heliosheath regions.  Gloeckler \& Geiss
(2002) derive the filtration factor for \HI\ by comparing H/He
measured in the pickup ion population with \HI/\HeI\ for the nearby
ISM.  Using their approach, Frisch and Slavin (2002) have shown that
Models 2 and 18 predict H filtration factors ($\sim$0.43) which yield
agreement between pickup ion $n$(\HI) values and the model
predictions for the solar location.  This is illustrated in Fig. 2.
%%\citep{FrischSlavin:2002}
%%\citep{RipkenFahr:1983,Izmodenovetal:1999}
%%\citep{CummingsStone:2002,MuellerZank:2002}

\begin{figure}[ht]
\vspace*{3.3in}
\includegraphics{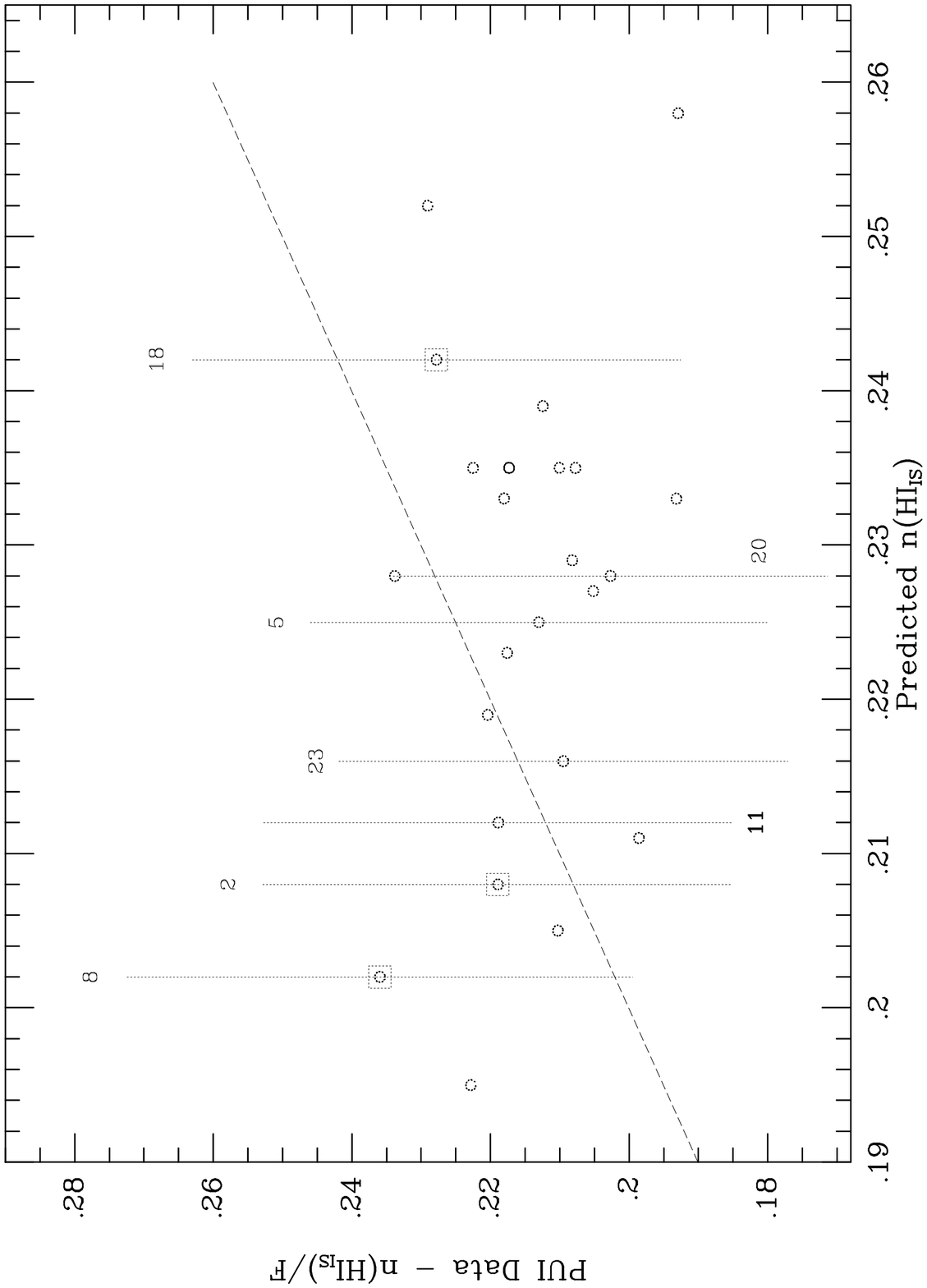}
\caption{ Comparison between \HI\ in the pickup ion population and
model predictions for $n$(\HI) for the 25 models of Slavin \& Frisch
(2002).  The ordinate in each plot is the interstellar $n$(\HI) value
that results after the pickup ion measurement of $n$(\HI) at the
termination shock is corrected for heliosheath filtration ($F$) using
\HI/\HeI\ at the solar location from the Slavin \& Frisch (2002) for
each model.  The abscissa for the plot is the predicted $n$(\HI) value
at solar location.  The viable models are shown with pickup ion data
uncertainties plotted, and the three best models based on Mg$^{\rm
+}$/Mg$^{\rm o}$ and C$^{\rm +}$/C$^{\rm +*}$ ratios towards
$\epsilon$ CMa have boxes around the data points.  The dotted line
illustrates perfect agreement between models and data.  Models 2 and
18 appear to provide the best match to the combined pickup ion and
$\epsilon$ CMa data when $n$(\HeI)=0.016$\pm$0.002 \cc\ in the LIC.  
The points for Models 16 and 17, which differ
only in the FUV radiation field, are superimposed on each other.
(Adopted from Frisch \& Slavin 2002.)
\label{Hfilt}}
\end{figure}

The gas-to-dust mass ratio (\Rgd) may be calculated from the radiative
transfer models, and the results suggest that interstellar gas and
dust are not well mixed in the LIC (Frisch \& Slavin 2002).  When
missing mass arguments and an assumed solar reference abundance are
applied to Model 18, a gas-to-dust mass ratio of $\sim$180 is found;
subsolar ISM reference abundances yield values a factor of $\sim$3 larger.
However, $in~situ$ dust observations of interstellar dust by Ulysses,
Galileo, and Cassini yield a gas-to-dust mass ratio of
$\sim$125$\pm$18 (Frisch et al. 1999; Landgraf et al. 2000; Frisch \&
Slavin 2002).  One explanation for the difference between these two
values for \Rgd\ would be poor gas-dust mixing as a superbubble shell
fragment sweeps up both gas and dust in the ISM.  Most of the mass of
the interstellar dust grains detected within the solar system is
carried by grains with masses $> 10^{-13}$ g (or radius $>$0.2
$\mu$m).  The gyro-radius for a grain with mass $10^{-13}$ g (radius
$\sim$0.2 $\mu$m) in a weak magnetic field ($\sim$3 $\mu$G) is
$\sim$0.3 pc (Gruen \& Landgraf 2000), which is greater than the
distance to the LIC edge in the upstream direction.  Thus massive dust
grains decouple from gas in the upstream LIC, and the \Rgd\ over the
small scales sampled by the in situ data differs from the value
averaged over a typical long interstellar sightlines.  The result is
that \Rgd\ may vary over sub-parsec length scales.  Other explanations,
such as poorly understood ISM abundances, are also possible.
%%\citep{Frischetal:1999,Landgrafetal:2000,FrischSlavin:2002}
%%\citep{GruenLandgraf:2000}
%%\citep{FrischSlavin:2002}

\section*{SOLAR MOTION AND ORIENTATION WITH RESPECT TO THE INTERSTELLAR MAGNETIC FIELD}

The Sun moves through space and has recently emerged from the interior
of the Local Bubble and entered an outflow of ISM from the
Scorpius-Centaurus Association (Fig. 1).  Hipparcos results yield a
solar motion in the local standard of rest of $\sim$13.4 \kms\ towards
the galactic coordinates $l$=28\deeg, $b$=+32\deeg\ (Dehnen \& Binney
1998).
%%\citep{DehnenBinney:1998}

The north ecliptic pole is directed towards $l$=96\deeg, $b$=30\deeg,
giving an ecliptic plane tilted by $\sim$60\deeg\ with respect to the
galactic plane.  Observations of the interstellar magnetic field
directly outside of the heliosphere have been elusive, but evidence
for a weak magnetic field ($<$3 $\mu$G) is given by starlight
polarization caused by interstellar dust grains aligned by nearby
magnetic fields (Tinbergen 1982, Frisch 1990).  The patch of ISM
causing this polarization is located towards $l$=0$\pm$20\deeg,
$b$=--20$\pm$25\deeg, and is close to the ecliptic plane towards the
nose direction of the heliosphere (see Fig. 3).  The polarization data
show a nearby magnetic field orientation which is approximately
parallel to the galactic plane and directed towards $l \sim $70\deeg.
The classical interstellar dust grains which cause the polarization
are charged, and pile up in the heliosheath regions as they are
deflected around the heliosphere (Frisch et al. 1999).  Although the
region of maximum polarization closely follows the ecliptic plane, it
also coincides with the CLIC LSR upstream direction (Fig. 3).
Observations by Voyager 1 and 2 observations of $\sim$3 kHz emission
in the outer heliosphere also indicate the interstellar magnetic field
direction is parallel to the galactic plane (Kurth \& Gurnett 2003).
The combined ecliptic tilt and interstellar magnetic field direction
produce a north/south asymmetry in the heliosphere configuration which
is visualized in Fig. 4.
%%\citep{Frisch:1990,Tinbergen:1982}
%%\citep{KurthGurnett:2003}
%%The direction of the electric vector polarization for light
%%

\section*{KINEMATICS AND TURBULENCE OF NEARBY ISM}

The LIC is one of a group of cloudlets which have a bulk flow motion
in the Local Standard of Rest (LSR) of --17$\pm$5 \kms, from the
upstream direction $l \sim$2.3\deeg, $b \sim$--5.2\deeg\ (Frisch et
al. 2002).  This group of cloudlets has been referred to as the
Cluster of Local Interstellar Clouds, or CLIC.  The dispersion of the
cloudlets about the bulk flow represents turbulence in the nearby ISM.
Models 2 and 18 indicate that $n$(\HI)=0.208--0.242 \cc\ in the LIC.
The hydrogen column density towards the nearest star, $\alpha$ Cen
(1.3 pc), has been reported as $N$(\HI)=0.41--1.04 10$^{18}$ \cmtwo\
(Linsky \& Wood 1996), yielding a filling factor of $f$=0.5--1 for ISM
in this sightline. Two cloudlets near the Sun have velocities consistent
with the absorption component towards $\alpha$ Cen, the ``G'' cloud
(Lallement \& Bertin 1992) and the Apex Cloud seen towards Aquila and
Ophiuchus (Frisch 2003).  The LSR upstream direction velocity vectors for the
G and Apex Clouds are, respectively, 
($V,l, b$)=(--17.7 \kms,350.5\deeg,9.2\deeg) and (--23.4 \kms,4.8\deeg,2.1\deeg).
For comparison the LSR upstream LIC motion is (--15.8 \kms,346.1\deeg,0.2\deeg). 
The Sun is located within a
few parsecs of the downstream edge of the bulk flow. The smaller
velocity of the LIC, located at the leading edge of the CLIC bulk
flow, suggests flow deceleration, while the dispersion of the CLIC
velocities reflects turbulence.  These flow parameters are derived
from observations of 96 interstellar absorption line components
(mainly Ca$^{\rm +}$ data) towards 60 stars which sample the ISM
within 30 parsecs of the Sun (Frisch, Grodnicki \& Welty 2002).  The LSR upstream
direction of the CLIC bulk flow is within $\sim$20\deeg\ of the
observed tangent point of Loop I, which has been formed by successive
epochs of star formation in the Scorpius-Centaurus Association over
the past 10$^7$ years.  The CLIC gas has been attributed to a
superbubble shell fragment associated with this star formation, with
possible implications for the chemical abundance pattern for this
nearest gas.  For such an origin, the CLIC would be composed of swept
up material consisting of ejecta from previous epochs of star
formation in the SCA, the ambient ISM, and recently ejected
nucleosynthetic products.  This scenario provides a hybrid method for
the dispersal of nucleosynthetic products -- e.g. cloud collisions and
turbulence within the outflow.
%%\citep{Frisch:2003}.
%%\citep{LinskyWood:1996}
%%\citep{Frischetal:2002}
%%\citep{Frischetal:2002}

\begin{figure}[ht]
\vspace*{3.2in}
\includegraphics{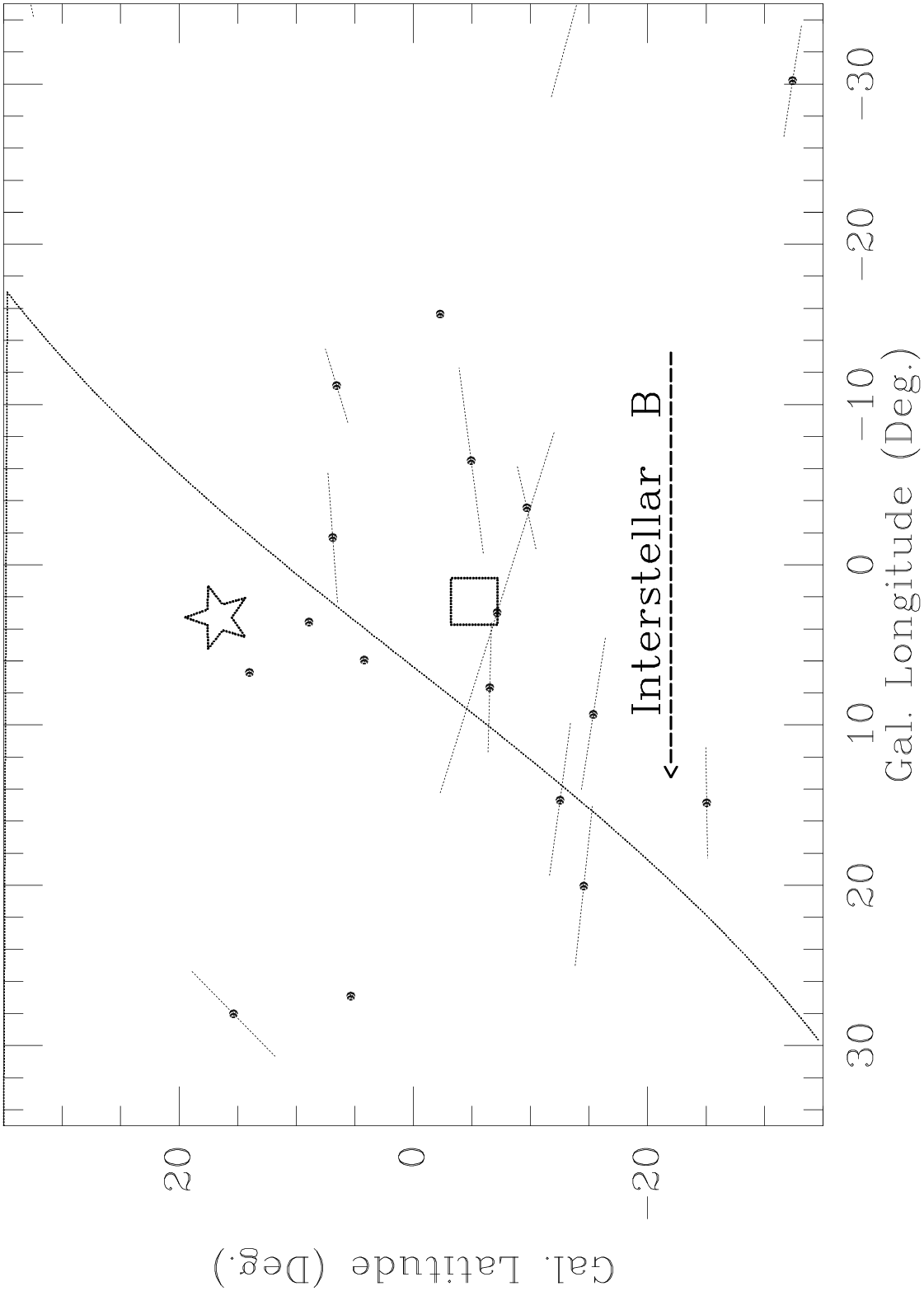}
\vspace*{-0.15in}
\caption{Plot of an indicator of the nearby interstellar magnetic field,
in galactic coordinates.
%The solid squares are the Voyager $\sim$3 kHz emission found by Kurth and Gurnett (2002).
The bars show the direction of the electric vector polarization, which
is parallel to the interstellar magnetic field direction, for several
nearby stars (Tinbergen 1982).  The arrow shows the likely direction
of the interstellar magnetic field near the Sun based on these data.
The curved line shows the ecliptic plane.  The region of maximum
polarization follows the ecliptic plane, but is also in the LSR
upstream CLIC direction (box).  A star is plotted at the heliosphere
nose direction in heliocentric coordinates.  The classical
interstellar dust grains which polarize optical radiation pile up in
the heliosheath regions as they are deflected around the heliosphere
(Frisch et al. 1999).
(Figure adopted from Frisch \& Slavin 2002.)
\label{Bfield}}
\end{figure}

\begin{figure}[t!]
\vspace*{3.5in}
%\special{psfile=Orion.eps voffset=0 hoffset=120 vscale=45 hscale=45 }
\caption{ Visualization of the asymmetric heliosphere caused by the
apparent tilt of the ecliptic plane with respect to the local
interstellar magnetic field (from Hanson et al. 2002).
The heliosphere morphology is based on an MHD simulation by 
Linde et al. (1998),
which includes the relative orientation and ram pressures of the
interstellar and solar wind magnetic fields.  Note the constellation
of Orion is viewed through the heliosphere tail.  The movie containing
this visualization can be downloaded from:
http://cs.indiana.edu/$\sim$soljourn.  (Used with permission.)
\label{fig:1}}
\end{figure}

%%\citep{Linde:1998,Linde:PhDthesis}
%%\citep{Hansonetal:2002b}

\section*{CONCLUSIONS}

Why study nearby ISM? Taking an anthropomorphic viewpoint, the ISM
dominates the interplanetary medium throughout most of the
heliosphere.  Heliosphere models show that heliosphere properties, and
by analogy exoplanetary system astrospheres, are dramatically affected
by the physical properties of the surrounding interstellar cloud.
From an astronomical viewpoint, the d$<$30 pc ISM is the best region
for studying the physical processes which operate in the ISM because
of the unusual range of data available for constraining radiative
transfer models.  These data include the ISM properties within the
solar system, which trace the neutral component.  Accurate ISM models
permit basic axioms underlying ISM studies to be tested, such as the
assumption that both gas and dust must be included when evaluating the
chemical composition of the ISM.

\section*{ACKNOWLEDGEMENTS}

The author would like to gratefully acknowledge research support from
NASA grants NAG5-6405, and NAG5-11005, and NAG5-8163.

%\section*{REFERENCES}

%%The references are formatted separately using \citep option in
%%agu2001.cls and agu.bst, with \def\ssr added to agu2001.

%%%%%%%%%%%%%%%%%%%%%%%%%%%%%%%%%%%%%%%%%%%%%%%%%%%%%%%%%%%%%%%%%%%%%%%%%%%%%%%%%%%%
%% Bookeeping stuff
%%
%%Corresponding author:  frisch@oddjob.uchicago.edu \\
$ ~~~ $

Manuscript received $3 ~ December ~ 2002$, revised $13 ~ February  ~ 2003$, accepted $13 ~ February  ~ 2003$.


\newpage

\begin{thebibliography}{42}
\expandafter\ifx\csname natexlab\endcsname\relax\def\natexlab#1{#1}\fi

\bibitem[{{\it {Adams} and {Frisch}\/}(1977)}]{AdamsFrisch:1977}
{Adams}, T.~F., and P.~C. {Frisch}, High-resolution observations of the
  \protect{{L}yman} alpha sky background, {\it \apj\/}, {\bf 212\/}, 300--308,
  1977.

\bibitem[{{\it {Allende Prieto} et~al.\/}(2001){\it {Allende Prieto},
  {Lambert}, and {Asplund}\/}}]{Prietoetal:2001}
{Allende Prieto}, C., D.~L. {Lambert}, and M.~{Asplund}, {The forbidden
  abundance of oxygen in the Sun}, {\it \apjl\/}, {\bf 556\/}, L63--L66, 2001.

\bibitem[{{\it {Andre} et~al.\/}(2002)}]{AndreHowketal:2002}
{Andre}, M., et~al., Oxygen gas phase abundance revisited, submitted to {\it \apj\/}, 2002.

\bibitem[{{\it Cummings et~al.\/}(2002){\it Cummings, Stone, and
  Steenberg\/}}]{CummingsStone:2002}
Cummings, A.~C., E.~C. Stone, and C.~D. Steenberg, {Composition of anomalous
  cosmic rays and other heliospheric ions}, {\it \apj\/}, {\it {\bf 578
  }\/}, 194--210, 2002.

\bibitem[{{\it {Dame} et~al.\/}(1987){\it {Dame}, {Ungerechts}, {Cohen}, {de
  Geus}, {Grenier}, {May}, {Murphy}, {Nyman}, and {Thaddeus}\/}}]{Dame:1987}
{Dame}, T.~M., H.~{Ungerechts}, R.~S. {Cohen}, E.~J. {de Geus}, I.~A.
  {Grenier}, J.~{May}, D.~C. {Murphy}, L.~A. {Nyman}, and P.~{Thaddeus}, A
  composite {CO} survey of the entire {M}ilky {W}ay, {\it \apj\/}, {\bf 322\/},
  706--720, 1987.

\bibitem[{{\it Dehnen and Binney\/}(1998)}]{DehnenBinney:1998}
Dehnen, W., and J.~J. Binney, Local stellar kinematics from {H}ipparcos data,
  {\it \mnras\/}, {\bf 298\/}, 387--394, 1998.

\bibitem[{{\it {Frisch}\/}(1990)}]{Frisch:1990}
{Frisch}, P.~C., Characteristics of the local interstellar medium, in {\it
  Physics of the Outer Heliosphere\/}, edited by S. Grzedzielski and D. E. Page,  pp. 19--22, Pergamon Press, Oxford,1990.

\bibitem[{{\it {Frisch}\/}(1993)}]{Frisch:1993a}
{Frisch}, P.~C., G-star astropauses - {A} test for interstellar pressure, {\it
  \apj\/}, {\bf 407\/}, 198--206, 1993.

\bibitem[{{\it {Frisch}\/}(1994)}]{Frisch:1994}
{Frisch}, P.~C., Morphology and ionization of the interstellar cloud
  surrounding the solar system, {\it {Science}\/}, {\bf 265\/}, 1423--1426, 1994.

\bibitem[{{\it Frisch\/}(1997)}]{Frisch:1997}
Frisch, P.~C., Journey of the {S}un, {\it http://xxx.lanl.gov/\/},
  astroph/9705231, 1997.

\bibitem[{{\it {Frisch} and {Slavin}\/}(2002)}]{FrischSlavin:2002}
{Frisch}, P.~C., and J.~D. {Slavin}, {Chemical composition and gas-to-dust mass
  ratio of the nearest interstellar matter}, submitted to {\it \apj, \/},
  2002.

\bibitem[{{\it {Frisch} et~al.\/}(2002){\it {Frisch}, {Grodnicki}, and
  {Welty}\/}}]{Frischetal:2002}
{Frisch}, P.~C., L.~{Grodnicki}, and D.~E. {Welty}, {The velocity distribution
  of the nearest interstellar gas}, {\it \apj\/}, {\bf 574\/}, 834--846, 2002.

\bibitem[{{\it {Frisch} et~al.\/}(2003){\it {Frisch}, {Mueller}, {Zank}, and
  {Lopate}\/}}]{Frischetal:2002b}
{Frisch}, P.~C., H.~R. {Mueller}, G.~P. {Zank}, and C.~{Lopate}, {\it {Galactic
  environment of the Sun and Stars: Interstellar and interplanetary
  material}\/}, in press, {Cambridge: Cambridge University Press}, 2003.

\bibitem[{{\it {Frisch} et~al.\/}(1999)}]{Frischetal:1999}
{Frisch}, P.~C.,  {Dorschner}, J. M. and {Geiss}, J.,  et~al., {Dust in the Local Interstellar Wind}, {\it \apj\/},
  {\bf 525\/}, 492--516, 1999.

\bibitem[{{\it {Gloeckler} and {Geiss}\/}(2002)}]{GloecklerGeiss:2002}
{Gloeckler}, G., and G.~{Geiss}, {Composition of the local interstellar medium
  as diagnosed with pickup ions}, {\it Adv. Space Res.}, this issue, 2003.

\bibitem[{{\it {Gruen} and {Landgraf}\/}(2000)}]{GruenLandgraf:2000}
{Gruen}, E., and M.~{Landgraf}, {Collisional consequences of big interstellar
  grains}, {\it \jgr\/}, {\bf 105\/}, 10291--10298, 2000.

\bibitem[{{\it {Gry} and {Jenkins}\/}(2001)}]{GryJenkins:2001}
{Gry}, C., and E.~B. {Jenkins}, Local clouds: Ionization, temperatures,
  electron densities and interfaces, from {GHRS} and {IMAPS} spectra of epsilon
  {C}anis {M}ajoris, {\it \aap\/}, {\bf 367\/}, 617--628, 2001.

\bibitem[{{\it Hanson et~al.\/}(2002){\it Hanson, Fu, Wernert, and
  Frisch\/}}]{Hansonetal:2002b}
Hanson, A.~J., C.-W. Fu, E.~A. Wernert, and P.~C. Frisch, Case study:
  Constructing the solar journey, {\it {Preprint}\/}, 2002.

\bibitem[{{\it Holweger\/}(2001)}]{Holweger:2001}
Holweger, H., Photospheric abundances: Problems, updates, implications, {\it
  http://xxx.lanl.gov/,astro-ph/0107426 \/}, 2001.

\bibitem[{{\it {Izmodenov} et~al.\/}(1999){\it {Izmodenov}, {Geiss},
  {Lallement}, {Gloeckler}, {Baranov}, and {Malama}\/}}]{Izmodenovetal:1999}
{Izmodenov}, V.~V., J.~{Geiss}, R.~{Lallement}, G.~{Gloeckler}, V.~B.
  {Baranov}, and Y.~G. {Malama}, Filtration of interstellar hydrogen in the
  two-shock heliospheric interface: Inferences on the local interstellar cloud
  electron density, {\it \jgr\/}, {\bf 104\/}, 4731--4742, 1999.

\bibitem[{{\it Kurth and Gurnett\/}(2003)}]{KurthGurnett:2003}
Kurth, W.~S., and D.~A. Gurnett, On the source location of low-frequency
  heliospheric radio emissions, submitted to {\it J. Geophys. Res.\/}, 2003.

\bibitem[{{\it {Lallement} et~al.\/}(1992){\it {Lallement}, and {Bertin} \/}}]{LallementBertin:1992}
{Lallement}, R., and P.~{Bertin},
  Northern-Hemisphere observations of nearby interstellar gas - Possible detection of the local cloud, {\it \aap\/}, {\bf 266\/}, 479--485, 1992.

\bibitem[{{\it {Landgraf} et~al.\/}(2000){\it {Landgraf}, {Baggaley}, {Gr{\"
  u}n}, {Kr{\" u}ger}, and {Linkert}\/}}]{Landgrafetal:2000}
{Landgraf}, M., W.~J. {Baggaley}, E.~{Gr{\" u}n}, H.~{Kr{\" u}ger}, and
  G.~{Linkert}, {Aspects of the mass distribution of interstellar dust grains
  in the solar system from in situ measurements}, {\it \jgr\/}, {\bf 105\/}, 10343--10352, 2000.

\bibitem[{{\it Linde\/}(1998)}]{Linde:PhDthesis}
Linde, T.~J., A three-dimensional adaptive multifluid {MHD} model of the
  heliosphere, Ph.D. thesis, Univ. of Michigan, Ann Arbor, 1998, {small \tt
  http://hpcc.engin.umich.edu/CFD/publications}.

\bibitem[{{\it Linde et~al.\/}(1998){\it Linde, Gombosi, Roe, Powell, and
  {DeZeeuw}\/}}]{Linde:1998}
Linde, T.~J., T.~I. Gombosi, P.~L. Roe, K.~G. Powell, and D.~L. {DeZeeuw},
  Heliosphere in the magnetized local interstellar medium: {R}esults of a
  three-dimensional {MHD} simulation, {\it \jgr\/}, {\bf 103\/}, 1889--1904,
  1998.

\bibitem[{{\it {Linsky} and {Wood}\/}(1996)}]{LinskyWood:1996}
{Linsky}, J.~L., and B.~E. {Wood}, The alpha {C}entauri line of sight: {D/H}
  ratio, physical properties of local interstellar gas, and measurement of
  heated hydrogen (the `hydrogen wall') near the heliopause, {\it \apj\/}, {\bf
  463\/}, 254--270, 1996.

\bibitem[{{\it {Lucke}\/}(1978)}]{Lucke:1978}
{Lucke}, P.~B., The distribution of color excesses and interstellar reddening
  material in the solar neighborhood, {\it \aap\/}, {\bf 64\/}, 367--377, 1978.

\bibitem[{{\it {M{\" u}ller} and {Zank}\/}(2002)}]{MuellerZank:2002}
{M{\" u}ller}, H., and G.~P. {Zank}, {Modeling heavy ions and atoms throughout
  the heliosphere}, in press, {\it "Solar Wind 10; AIP Conference Proceedings"\/}, 2002.

\bibitem[{{\it {Marsh} and {Svensmark}\/}(2000)}]{MarshSvensmark:2000}
{Marsh}, N.~D., and H.~{Svensmark}, {Low cloud properties influenced by cosmic
  rays}, {\it Physical Review Letters\/}, {\bf 85\/}, 5004-5007, 2000.

\bibitem[{{\it {Mueller} et~al.\/}(2001){\it {Mueller}, {Zank}, and
  {Frisch}\/}}]{Muelleretal:2001}
{Mueller}, H.~R., G.~P. {Zank}, and P.~C. {Frisch}, Effect of different
  possible interstellar environments on the heliosphere: A numerical study, in
  {\it "The Outer Heliosphere: The Next Frontiers"\/}, edited by K.~{Scherer},
  COSPAR Colloquiua Series, 11, 329--333, 2001.

\bibitem[{{\it {Mueller} et~al.\/}(2002){\it {Mueller}, {Zank}, and
  {Frisch}\/}}]{Muelleretal:2002}
{Mueller}, H.~R., G.~P. {Zank}, and P.~C. {Frisch}, Heliospheric response to
  different possible interstellar environments, {\it {\it in preparation}\/},
  2002.

\bibitem[{{\it {P.~C.~Frisch}\/}(2003)}]{Frisch:2003}
{P.~C.~Frisch}, {Local interstellar matter: The Apex Cloud}, submitted to {\it \apj\/}, 2003.

\bibitem[{{\it {Ripken} and {Fahr}\/}(1983)}]{RipkenFahr:1983}
{Ripken}, H.~W., and H.~J. {Fahr}, {Modification of the local interstellar gas
  properties in the heliospheric interface}, {\it \aap\/}, {\bf 122\/}, 181--192, 1983.

\bibitem[{{\it {Roble}\/}(1991)}]{Roble:1991}
{Roble}, R.~G., {On modeling component processes in the earth's global electric
  circuit}, {\it Journal of Atmospheric and Terrestrial Physics\/}, {\bf 53\/}, 831-847, 1991.

\bibitem[{{\it {Slavin} and {Frisch}\/}(2002)}]{SlavinFrisch:2002}
{Slavin}, J.~D., and P.~C. {Frisch}, {The Ionization of Nearby Interstellar
  Gas}, {\it \apj\/}, {\bf 565\/}, 364--379, 2002.

\bibitem[{{\it Tinbergen\/}(1982)}]{Tinbergen:1982}
Tinbergen, J., Interstellar polarization in the immediate solar neighborhoud,
  {\it \aap\/}, {\bf 105\/}, 53--64, 1982.

\bibitem[{{\it {Tinsley}\/}(2000)}]{Tinsley:2000}
{Tinsley}, B.~A., {Influence of solar wind on the global electric circuit, and
  inferred effects on cloud microphysics, temperature, and dynamics in the
  Troposphere}, {\it Space Science Reviews\/}, {\bf 94\/}, 231--258, 2000.

\bibitem[{{\it Witte et~al.\/}(1996){\it Witte, Banaszkiewicz, and
  Rosenbauer\/}}]{Witte:1996}
Witte, M., M.~Banaszkiewicz, and H.~Rosenbauer, Recent results on the
  parameters of the interstellar helium from the {ULYSSES/GAS} experiment, {\it
  \ssr\/}, {\bf 78\/}, 289--2396, 1996.

\bibitem[{{\it {Wood} et~al.\/}(2001){\it {Wood}, {Linsky}, {M{\" u}ller}, and
  {Zank}\/}}]{Woodetal:2001}
{Wood}, B., J.~{Linsky}, H.~{M{\" u}ller}, and G.~{Zank}, {Observational
  estimates for the mass-loss rates of alpha; Centauri and Proxima Centauri
  using Hubble Space Telescope Ly-alpha spectra}, {\it \apjl\/}, {\it 547 \/}, L49--L52, 2001.

\bibitem[{{\it {Wood} et~al.\/}(2002){\it {Wood}, {M{\" u}ller}, {Zank}, and
  {Linsky}\/}}]{Woodetal:2002}
{Wood}, B., H.~{M{\" u}ller}, G.~{Zank}, and J.~{Linsky}, {Measured mass-.oss
  rates of solar-like stars as a function of age and activity}, {\it \apjl\/},
  {\bf 574\/}, 412--425, 2002.

\bibitem[{{\it {Zank} and {Frisch}\/}(1999)}]{ZankFrisch:1999}
{Zank}, G.~P., and P.~C. {Frisch}, Consequences of a change in the {G}alactic
  environment of the {S}un, {\it \apj\/}, {\bf 518\/}, 965--973, 1999.

\end{thebibliography}
\end{document}